\documentclass{WileyMSP-template}

\usepackage{graphicx}%
\usepackage[margin=1cm,font=small,labelfont=bf,format=plain,labelsep=period, justification=justified,figureposition=below, tableposition=top, skip=5pt]{caption}
\usepackage{multirow}%
\usepackage{amsmath,amssymb,amsfonts}%
\usepackage{mathtools}
\usepackage{amsthm}%
\usepackage{mathrsfs}%
\usepackage[title]{appendix}%
\usepackage{xcolor}%
\usepackage{textcomp}%
\usepackage{manyfoot}%
\usepackage{booktabs}%
\usepackage{algorithm}%
\usepackage{algorithmicx}%
\usepackage{algpseudocode}%
\usepackage{listings}%
\usepackage{geometry}
\begin{document}

\pagestyle{fancy}
\rhead{\includegraphics[width=2.5cm]{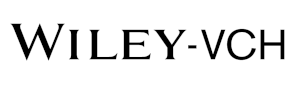}}

\title{Low-Loss and Low-Power Silicon Ring Based WDM 32$\times$100 GHz Filter Enabled by a Novel Bend Design}

\maketitle


\author{Qingzhong Deng*}
\author{Ahmed H. El-Saeed}
\author{Alaa Elshazly}
\author{Guy Lepage}
\author{Chiara Marchese}
\author{Pieter Neutens}
\author{Hakim Kobbi}
\author{Rafal Magdziak}
\author{Jeroen De Coster}
\author{Javad Rahimi Vaskasi}
\author{Minkyu Kim}
\author{Yeyu Tong}
\author{Neha Singh}
\author{Marko Ersek Filipcic}
\author{Pol Van Dorpe}
\author{Kristof Croes}
\author{Maumita Chakrabarti}
\author{Dimitrios Velenis}
\author{Peter De Heyn}
\author{Peter Verheyen}
\author{Philippe Absil}
\author{Filippo Ferraro}
\author{Yoojin Ban}
\author{Joris Van Campenhout}

\begin{affiliations}
imec, Leuven, B-3001 Belgium\\
Email Address: qingzhong.deng@imec.be

\end{affiliations}


\begin{abstract}

Ring resonators are essential components in silicon photonics with a wide range of applications, such as modulating, adding, and dropping signals in an optical communication system with wavelength division multiplexing (WDM).
However, the performance of conventional circular rings is limited by several trade-offs, such as free spectral range (FSR), ring roundtrip loss, ring-bus coupling, and resonance tuning efficiency.
To overcome these limitations, we propose a third order polynomial interconnected circular (TOPIC) bend to revolutionize the ring designs fundamentally.
The TOPIC bend has a unique feature of continuous curvature and curvature derivative, which is theoretically derived to be essential for waveguide loss optimization.
This unique feature allows the TOPIC bend to reduce the loss by more than 22 times compared to a circular bend, and more than 14 times compared to the existing low-loss bend designs including the widely used Euler bend as experimentally observed in a bend comparison.
The TOPIC bend can be easily configured to propagate the light as a whispering gallery mode, which is exploited to integrate a doped silicon heater into the waveguide.
With the TOPIC bend, the silicon ring resonators demonstrated here have achieved three records to the best of our knowledge: the smallest radius (0.7 $\mathrm{\mu m}$) for silicon rings resonating with single guided mode, the lowest thermal tuning power (5.85 mW/$\mathrm{\pi}$) for silicon rings with FSR$\geq$3.2 THz, and the first silicon ring-based WDM 32$\mathrm{\times}$100 GHz filter.
The filter has doubled the channel amount compared to the state of the art, and meanwhile achieved low insertion loss (1.91$\pm$0.28 dB) and low tuning power (283 GHz/mW).
Moreover, the TOPIC bend is not limited to ring applications,
it can also be used to create bends with an arbitrary angle, with the advantages of ultra-compact radius and heater integration, which are expected to replace all circular bends in integrated photonics, greatly reducing system size and power consumption.

\end{abstract}

\section{Introduction}
\begin{figure}[!b]%
    \centering
    \includegraphics{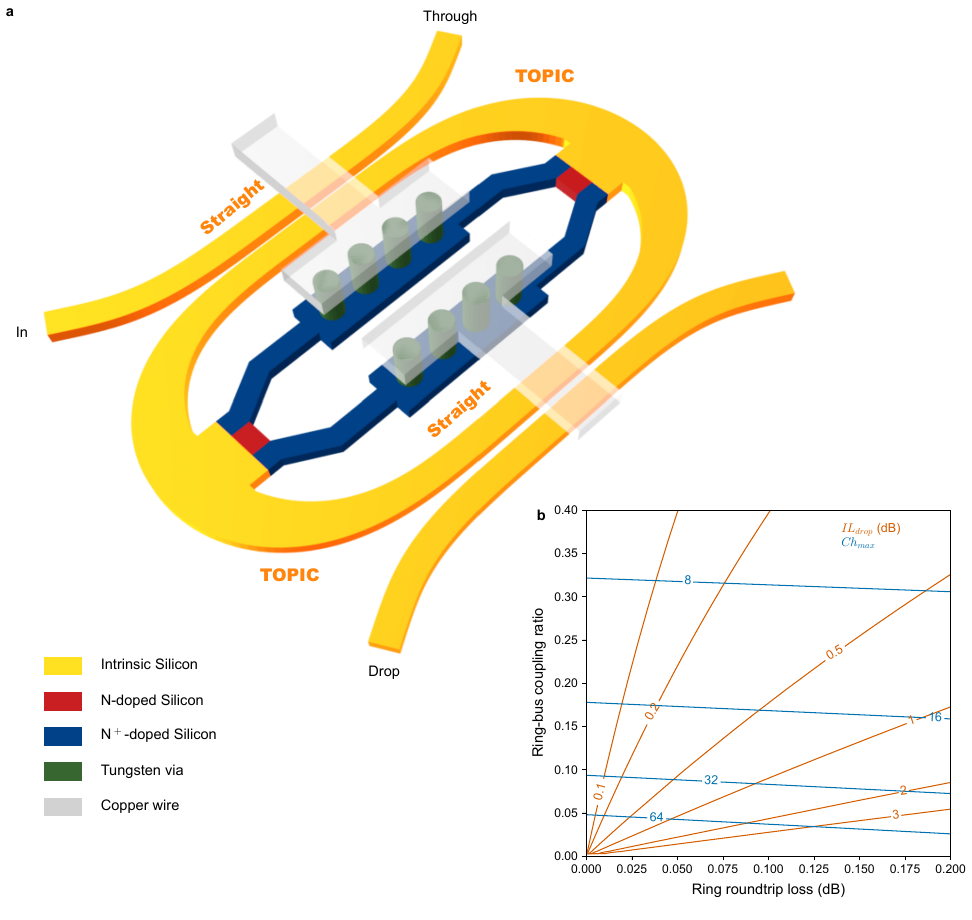}
    \caption{\textbf{The schematic of the WGM-based add-drop ring resonator with doped Si heaters.}
    \textbf{a,} The schematic of the add-drop ring constructed with the proposed TOPIC bend.
    \textbf{b,} The impact of the ring-bus coupling and ring roundtrip loss to ring based WDM filters.
    $IL_{drop}$ is the insertion loss at the ring dropping central wavelength, and $Ch_{max}$ is the maximum WDM channel amount determined by the full width at half maximum (FWHM) in the drop spectrum (see Note1, Supplementary Information, for the theoretical modelling of the add-drop ring resonator).}
    \label{fig_scheme_3d}
\end{figure}

Silicon photonics, a groundbreaking technology at the intersection of electronics and photonics, has emerged as a transformative force in the relentless pursuit for faster, more energy-efficient, and compact data transmission solutions~\cite{ReviewSiliconPhotonics_JOLT2021a}.
Data presented by electrical signals are modulated onto a light beam and added to a bus waveguide for transmission.
Multiple light beams with different wavelengths can propagate in the same bus waveguide to transfer data in parallel, well-known as wavelength division multiplexing (WDM).
At the receiver side, the light with a specific wavelength is dropped from the bus waveguide for signal detection~\cite{450GbsAll_2021a}.
To modulate, add, and drop signals at a filtered light wavelength, 
silicon ring resonators have been extensively investigated as an elegant approach owing to the compact size and promising performance~\cite{IntegratingPhotonicsSilicon_N2018a}.
Despite their potential, existing silicon ring-based WDM filters are limited to a maximum of 16 channels with high insertion loss~\cite{SiliconBasedChip_P2023a, SiMicroRing_OE2011a, Novel16Channel_2022a}.
The maximum WDM channel amount is bounded by the free spectral range (FSR) of the ring.
A large FSR can be achieved by a compact ring with a small bend radius, but a small bend radius in general will cause a high ring roundtrip loss that will dramatically degrade the ring performance, such as the dropping insertion loss (Fig.~\ref{fig_scheme_3d}b).
Moreover, a certain ring-bus coupling ratio is required to achieve a specific filtering spectrum~\cite{ResponseShapingSilicon_NP2018a}.
Last but not least, the operating wavelengths of a silicon ring are very sensitive to fabrication variations and ambient temperature such that energy-efficient wavelength tuning is always expected in practical applications.
In short, to unleash the full potential of silicon ring resonators, a large ring FSR, low ring roundtrip loss, scalable ring-bus coupling, and energy-efficient resonance tuning are simultaneously required.
However, these four requirements have not been fulfilled yet at the same time.
Rings consisting of circular bends with small radii were explored for a large FSR~\cite{ExtremeMiniaturizationSilicon_IPJ2010a,SiliconMicroringResonators_OE2008a,SubmicronResonatorBased_OE2019a,TunableSiliconMicroring_APL2006a,UltracompactSoiMicroring_IPTL2009a}.
These rings are quite lossy and have limited coupling ratio options, which can be mitigated by racetrack rings with smooth bend designs~\cite{EnhancedSiliconRing_IPTL2021a,MicroringResonatorDesign_2015a,LowLosswideFsr_2023a}. 
To tune the working wavelengths of these rings, heaters have to be placed micro-meters away from the ring to avoid high ring roundtrip loss~\cite{LowLosswideFsr_2023a,UltrabroadBandwidthArbitrary_NP2010a}, which results in limited tuning efficiency.
The heaters can be integrated into the ring for high tuning efficiency when whispering gallery mode (WGM) is excited in the ring waveguide~\cite{IntegratedMicroringTuning_2013a,AdiabaticResonantMicroring_2013a,AdiabaticMicroringResonators_OL2010a,AdiabaticResonantMicrorings_2COLAEAQEALSCC2V12009a}.
However, these existing WGM-based rings will be lossy when extended to be racetrack rings for a scalable coupling ratio.

In this paper, the ring resonators are revolutionized due to the third order polynomial interconnected circular (TOPIC) bend design which is proposed based on rigorous derivations starting from the general loss optimization in waveguides.
Compared to the existing low-loss bend designs~\cite{LowLossUltracompact_OE2023a, LowLossModified_OE2022a,NeuralAdjointMethod_OE2021a,OptimalBezierCurve_OL2021a,RobustEfficientMicrometre_NP2021a,UltrahighQSilicon_PR2020a,InverseDesignedPhotonic_JOLT2020a,LowLossWaveguide_JOLT2020a,AnalysisSiliconNitride_OE2019a,ArbitrarilyRoutedMode_NC2019a,UltraSharpMultimode_LPR2019a,UniversalDesignWaveguide_JOLT2019a,GeneralSBend_JOPCS2019a,LowLossLow_OE2018a,TopologyOptimalDesign_IEE2018a,VerySharpAdiabatic_OL2018a,PhotonicWeldingPoints_N2018a,LowLossCompact_OE2017a,AdiabaticallyBentWaveguides_IPTL2016a,MicroringResonatorDesign_2015a,OptimizationAdiabaticMicroring_AO2015a,LowLossHigh_OL2014a,AdiabaticallyWidenedSilicon_OE2014a,DramaticSizeReduction_OE2013a,AdiabaticThermoOptic_OL2013a,GeneralDesignAlgorithm_OE2012a,ChipTransformationOptics_NC2012a,AdiabaticMicroringModulators_OE2012a,CompactSingleMode_IPJ2011a,GeneralScalingRule_JOLT2011a,AdiabaticMicroringResonators_OL2010a,DesignLowLoss_OAQE2008a,IdealBendContour_IPTL2007a,ComputingOptimalWaveguide_JOLT2007a,RadiationModesRoughness_IJOSTIQE2006a,DesignCurvedWaveguides_JA2003a,MeasurementModeField_JOLT1997a,BendingLossReduction_JOLT1995a,NewGeneralApproach_JOLT1995a,NormalizedApproachDesign_JOLT1993a},
the TOPIC bend has a smoother transition to straight waveguides with a unique feature of continuous curvature and continuous curvature derivative everywhere, which minimizes the mode transition loss.
Moreover, the TOPIC bend has one adjustable design parameter that acts as an elegant approach to trade off the mode transition loss and the sidewall scattering loss.
When this parameter is adjusted for the inner and outer boundary in a TOPIC bend separately, WGM can be excited to reduce the loss further and seamlessly integrate the heaters within the optical waveguides.
We have presented some preliminary results in recent conferences~\cite{LowLossWaveguide_2ISPCS2023a, 32x100GhzWdm_OFCCO22024a},
while the comprehensive analysis and better devices performance are discussed here in detail.
The proposed TOPIC bend is experimentally shown to substantially reduce the loss to a mere $0.017\pm0.005$ dB, as compared to $0.378\pm0.026$ dB, $0.293\pm0.030$ dB , and $0.242\pm0.006$ dB in a circular bend, an Euler bend ~\cite{DramaticSizeReduction_OE2013a}, and a traditional WGM bend ~\cite{AdiabaticMicroringResonators_OL2010a} respectively.
Deploying this bend design to silicon ring resonators, we have set three records to our best knowledge:
the smallest radius (0.7 $\mathrm{\mu m}$) for silicon rings resonating with a single guided mode,
the lowest thermal tuning power (5.85 mW/$\mathrm{\pi}$) for silicon rings with an FSR$\geq$3.2 THz,
and the first silicon ring-based WDM 32$\mathrm{\times}$100 GHz filter.


\section{Third order polynomial interconnected circular (TOPIC) bend}

\begin{figure}[!b]%
    \centering
    \includegraphics{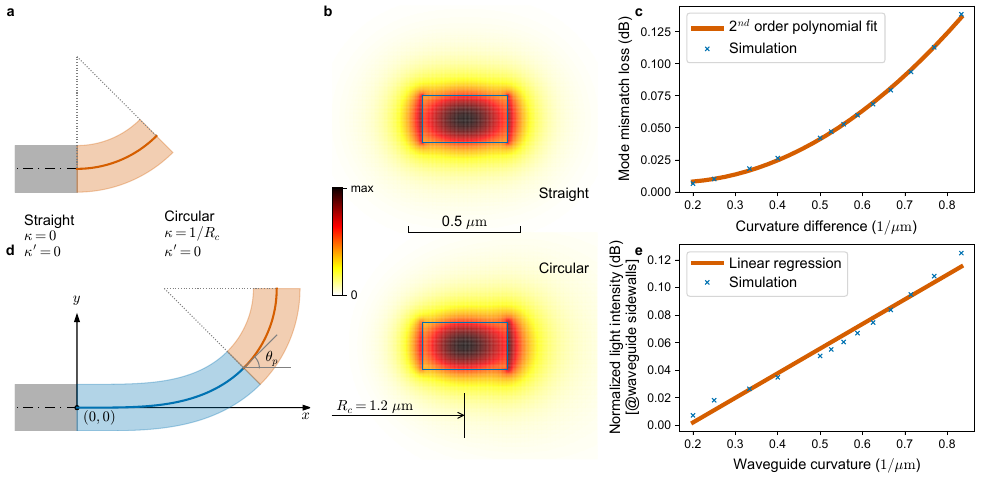}
    \caption{\textbf{The optical mode distortion in waveguide bends.}
    \textbf{a,d,} The schematic of the direct connection (\textbf{a}) and the connection with a smooth transition (\textbf{d}) between a straight waveguide and a circular bend.
    \textbf{b,} The optical field distributions ($|\mathbf{E}|$) of the fundamental TE mode in a straight and a circular waveguide.
    \textbf{c,} The mode transition loss between circular bends and the straight waveguide (see Note2, Supplementary Information, for the mode transition loss calculation based on eigenmode simulations).
    \textbf{e,} The average light intensity at the waveguide left and right sidewalls of circular bends.
    For all simulations in this paper, if not otherwise specified, the waveguide material stacks are SOI with silicon oxide as the top cladding, the analysis is carried out for fundamental TE mode, the Si thickness is 210 nm, the waveguide width is 380 nm, and the working wavelength is 1310 nm.
    }
    \label{fig_mode_overlap}
\end{figure}

A waveguide bend, usually circular, is one of the most fundamental components in photonic integrated circuits.
It has four sources of loss in general: material absorption, radiation due to bending, material scattering, and mode transition when connected to another waveguide (Fig.~\ref{fig_mode_overlap}a).
In silicon photonics, material absorption loss is negligible, and so is the radiation loss except for extremely small radii~\cite{AnalysisCurvedOptical_IJOQE1975a} that are not used in practice.
Material scattering loss mainly stems from the sidewall roughness caused by the etching process.
Hence, sidewall scattering and mode transition losses are the two main loss factors in silicon photonic bends.
When a waveguide is bent, the light propagating inside it undergoes distortion and shifts towards the outer boundary of the bend, as illustrated in Fig.~\ref{fig_mode_overlap}b.
The degree of distortion increases with the bending curvature.
The mode transition loss can be approximated as a second order polynomial function with respect to the waveguide curvature difference as shown in Fig.~\ref{fig_mode_overlap}c.
On the other hand, sidewall scattering loss in a waveguide is proportional to the sidewall light intensity whose average increases linearly with the bending curvature as shown in Fig.~\ref{fig_mode_overlap}e.
Based on these facts, the optical loss in a constant-width waveguide can be expressed as
\begin{equation}
    Loss=\int_{\kappa_{1}}^{\kappa_{2}}\left(A+B\kappa+C\frac{\mathrm{d}\kappa}{\mathrm{d}s}+D\left(\frac{\mathrm{d}\kappa}{\mathrm{d}s}\right)^2\right)\frac{\mathrm{d}s}{\mathrm{d}\kappa}d\kappa
    \label{eq_loss_func}
\end{equation}
where $s$ is the distance to the waveguide starting point and the curvature $\kappa$ is a function of $s$.
Both $s$ and $\kappa$ are defined along the waveguide central baseline.
Constant $\kappa_{1}$ and $\kappa_{2}$ are the curvatures at the waveguide start and end respectively.
Constant $A$ and $B$ denote the sidewall scattering loss, while constant $C$ and $D$ represent the mode transition loss.
To design a low-loss waveguide means, mathematically, to find an expression of $\kappa$ with respect to $s$ that minimizes $Loss$.
This optimization problem can be solved by Euler-Lagrange equation with the solution expressed as (see Note3, Supplementary Information, for the derivation details)
\begin{equation} 
    \kappa'' = \mathrm{Constant}
    \label{eq_optimal_solution}.
\end{equation}
This solution indicates that the optical loss is minimized when the second-order curvature derivative ($\kappa''\stackrel{\text{def}}{=}\mathrm{d}^{2}\kappa/\mathrm{d}s^{2}$) has no variation along the waveguide.
The physical meanings of this optimal solution are that the curvature $\kappa$ and the curvature derivative ($\kappa'\stackrel{\text{def}}{=}\mathrm{d}\kappa/\mathrm{d}s$) should be continuous along the waveguide, and the curvature derivative should vary linearly with respect to the waveguide length.
Both straight and circular waveguides are typical optimal designs following Eq.~\ref{eq_optimal_solution}, that explains why extra loss is always observed when another type of waveguide segment is inserted between two identical straight waveguides or two identical circular waveguides.
Eq.~\ref{eq_optimal_solution} can be used to design low-loss transitions between two different waveguides with ($\kappa=\kappa_1$, $\kappa'=\kappa'_1$) and ($\kappa=\kappa_2$, $\kappa'=\kappa'_2$) respectively when

\begin{equation} 
    (\kappa_2-\kappa_1)(\kappa'_2+\kappa'_1)\neq 0
    \label{eq_validation_condition}
\end{equation}
where the subscripts 1 and 2 denote the two waveguides.
However, the transition between a straight and a circular waveguide (Fig.~\ref{fig_mode_overlap}d) does not meet the condition given by Eq.~\ref{eq_validation_condition} because $\kappa'_1=\kappa'_2=0$.
Thus, we relaxed the optimal solution from ``no variation in $\kappa''$'' to ``no abrupt variation in $\kappa''$''.
In particular, a linear variation of $\kappa''$ with respect to $s$ is used as the relaxed optimal solution that results in a straight-circular transition with a third order polynomial curvature 
\begin{equation}
    \kappa=\dfrac{3R_{c}\theta_{p} s^2-s^3}{4R_{c}^{4}\theta_{p}^{3}}
    \label{eq_top_curvature}.
\end{equation}
where $R_{c}$ is the radius of the circular bend and $\theta_{p}$ is the total angle of the transition as marked in Fig.~\ref{fig_mode_overlap}d.
This equation can generate a low-loss transition waveguide, connecting a straight waveguide and a circular bend with a known radius $R_{c}$, with a total length of $2R_{c}\theta_{p}$ when a transition angle $\theta_{p}$ is specified.

\begin{figure}[!b]
    \centerline{\includegraphics{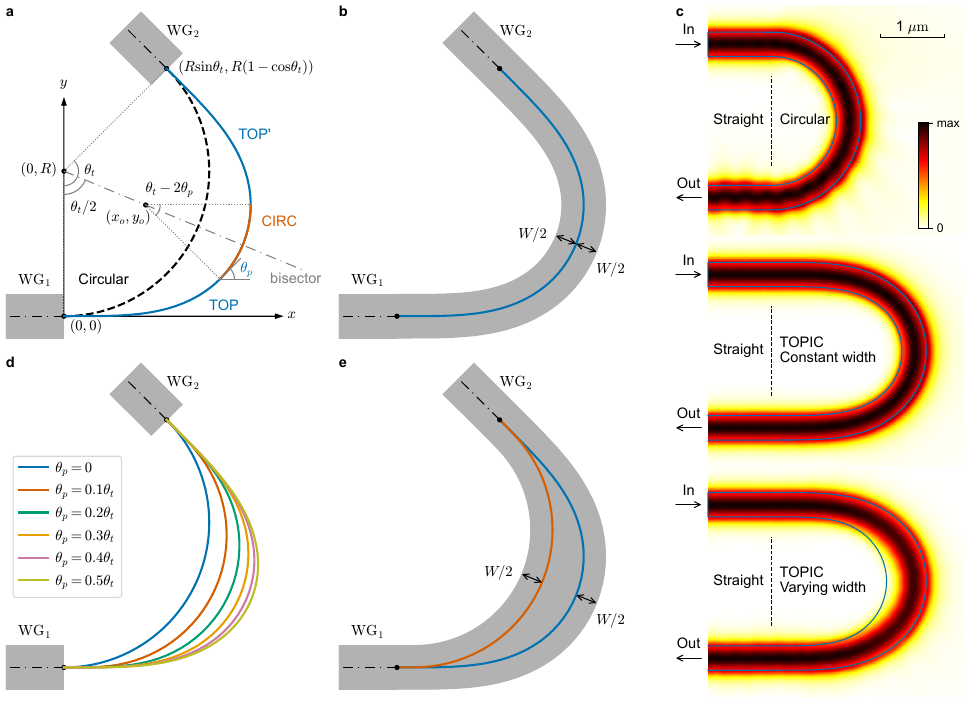}}
    \caption{\textbf{The schematics of TOPIC bend design.}
    \textbf{a,d,} The schematic of TOPIC curve with comparison to the traditional circular curve (\textbf{a}) and the impact of $\theta_p$ (\textbf{d}).
    \textbf{b,e,} The schematic of the TOPIC bend with constant (\textbf{b}) and varying (\textbf{e}) width.
    The constant-width bend is generated by constructing the inner and outer boundary as parallel curves of the same TOPIC baseline, while the varying-width bend is generated by constructing the inner and outer boundary as parallel curves of two TOPIC baselines with different $\theta_p$.
    \textbf{c,} The optical field evolutions ($|\mathbf{E}|$) in different bends with $R=1.2\ \mathrm{\mu m}$.
    All optical field evolutions in this paper are simulated by COMSOL with 3D full-vector finite element method, and the plotted field evolutions are sliced at the center of the waveguides along the thickness direction.
    }
    \label{fig_bend_scheme}
\end{figure}

Usually, both the start and the end of a circular bend are connected to straight waveguides.
To lower the loss, we propose to replace the circular bend by a TOPIC bend, as shown by the central baselines in Fig.~\ref{fig_bend_scheme}a.
The radius and the angle of the original circular bend are annotated as $R$ and $\theta_t$, respectively.
The TOPIC bend, consisting of 3 segments (TOP, CIRC and TOP'), is mirror-symmetric with respect to the bisector of $\theta_t$ so that only TOP and CIRC need to be defined.
TOP is a bend generated by Eq.~\ref{eq_top_curvature}, while CIRC is a circular bend with angle $\theta_t-2\theta_p$ and radius $R_c$ solved from
\begin{equation}
    \begin{dcases}
    \displaystyle &x_o\cos{\dfrac{\theta_t}{2}}+(y_o-R)\sin{\dfrac{\theta_t}{2}}=0\\ &x_o=\int_{0}^{2R_c\theta_p}\cos{\theta} \mathrm{d}s - R_c \sin\theta_p \\ &y_o=\int_{0}^{2R_c\theta_p}\sin{\theta} \mathrm{d}s + R_c \cos\theta_p \\ &\theta=\int_{0}^{s}\kappa \mathrm{d}s=\dfrac{4R_c\theta_p s^3-s^4}{16R_c^4{\theta_p}^3}\\
    \end{dcases}.
    \label{eq_for_rc}
\end{equation}
The physical meaning of Eq.~\ref{eq_for_rc} is that the osculating circle center at the end of TOP must be on the bisector of $\theta_t$ so that the circular segment CIRC can be connected between TOP and TOP'.
$\theta$ in this equation is the tangential angle along the bend that can be used to calculate the coordinates on the TOP segment

\begin{equation} 
    \left(x(l),y(l)\right)=\int_{0}^{l}(\cos{\theta}, sin{\theta})ds,\ \ \mathrm{for}\ 0\leq l\leq 2R_c\theta_p.
    \label{eq_for_xy}
\end{equation}

With these equations, the generation of a TOPIC baseline can be summarized as:
($\mathrm{I}$) Set a value for $\theta_p$ in the range of $0\sim \theta_t/2$;
($\mathrm{II}$) Calculate $R_c$ with Eq.~\ref{eq_for_rc};
($\mathrm{III}$) Generate a TOP segment by Eq.\ref{eq_for_xy};
($\mathrm{IV}$) Generate a CIRC segment as a circular bend, starting from the end of TOP, with a center $(x_o,y_o)$, a radius $R_c$, and an angle $\theta_t - 2\theta_p$;
($\mathrm{V}$) Generate the TOP' segment by mirroring TOP with respect to the bisector of $\theta_t$.
A curve generated by this method has continuous curvature and curvature derivative everywhere as required by the waveguide loss optimization.
Moreover, the curve shape can be manipulated by varying $\theta_p$ as shown in Fig.~\ref{fig_bend_scheme}d.
The total length of the TOP segment is $2R_{c}\theta_{p}$.
As $\theta_{p}$ increases, the baseline becomes longer and the maximum curvature along the baseline becomes larger, which leads to higher sidewall scattering loss, but also slows down the curvature variation, which lowers the mode transition loss.
Therefore, $\theta_{p}$ can be used as a tuning parameter to balance these two sources of loss to minimize the total loss.

\begin{figure}[!b]
    \centerline{\includegraphics{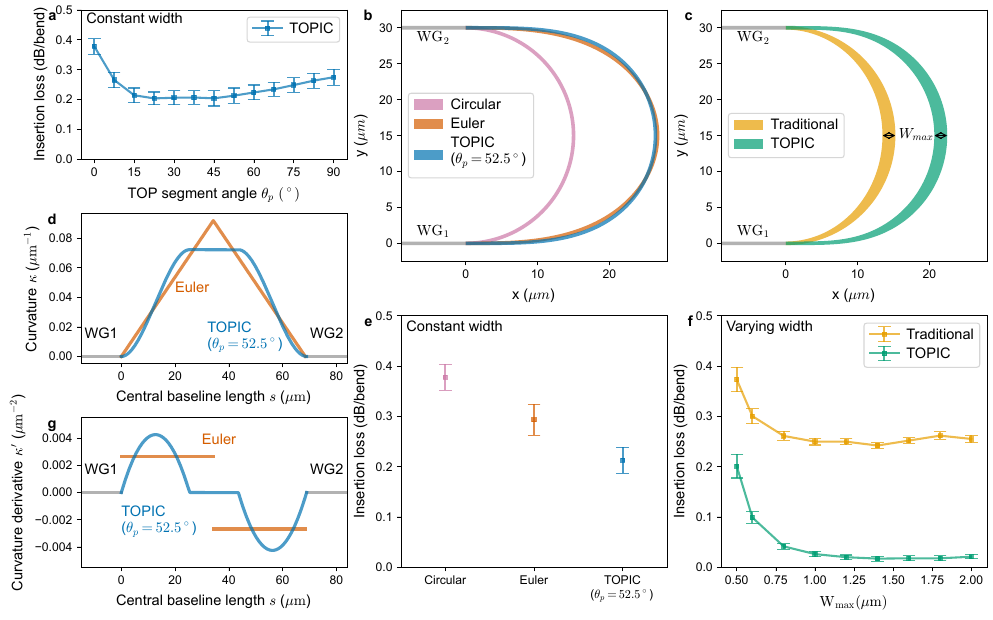}}
    \caption{
    \textbf{The measured insertion loss of different $180^\circ$ waveguide bends.}
    \textbf{a,} The measured loss of constant-width TOPIC bends with different $\theta_p$.
    \textbf{b,c,e,f,} The schematic (\textbf{b, c}) and the corresponding measured loss (\textbf{e, f}) of the waveguide bends with constant width (\textbf{b,e}) and varying width (\textbf{c, f}).
    \textbf{d,g,} The comparison of curvature (\textbf{d}) and curvature derivative (\textbf{g}) between the Euler and TOPIC bends with constant width.
    The waveguide material stacks are silicon nitride (SiN) surrounded by silicon oxide.
    The working wavelength is 638 nm.
    The SiN thickness is 150 nm.
    All the waveguide width, except the regions with varying width, is 500 nm.
    The bend radius $R=15\ \mathrm{\mu m}$.
    $W_{max}$ is the maximum waveguide width along a waveguide bend with varying widths.
    In traditional varying-width waveguides, the outer boundary is fixed as a circular curve while the inner boundary is an oval curve whose semi-minor axis is adjusted to get a certain $W_{max}$.
    In varying-width TOPIC bends, $\theta_p$ of the outer boundary $\theta_{p,o}$ is fixed to be $30^\circ$, and $\theta_p$ of the inner boundary $\theta_{p,i}$ is adjusted to get a certain $W_{max}$.
    The losses are measured using the cut-back method, where varying numbers of identical bends are cascaded for transmission measurements. The loss is then quantified through linear regression analysis of the transmission data.
    The plotted loss data are the mean and standard deviation of the measurements over all dies in a 200 mm wafer fabricated by imec pix4life platform.
    }
    \label{fig_bend_loss}
\end{figure}

When a TOPIC baseline is generated, it is straight-forward to get a constant-width TOPIC bend by constructing the waveguide inner and outer boundaries as parallel curves of the baseline as shown in Fig.~\ref{fig_bend_scheme}b.
Furthermore, a varying-width TOPIC bend can be easily generated by constructing the waveguide inner and outer boundaries as parallel curves of two baselines with different $\theta_p$ as depicted in Fig.~\ref{fig_bend_scheme}e.
The TOPIC bends can significantly reduce the mode transition loss compared to a circular bend, as verified by the optical field evolutions in Fig.~\ref{fig_bend_scheme}c.
This figure also indicates that the varying-width TOPIC bend can dramatically reduce the light intensity at the waveguide sidewalls, especially at the inner side, which is expected to greatly reduce the sidewall scattering loss with the existence of sidewall roughness.
To verify these findings, different types of silicon nitride bends are fabricated to connect two straight waveguides with fixed relative positions, as marked by $\mathrm{WG_{1}}$ and $\mathrm{WG_{2}}$ in Fig.~\ref{fig_bend_loss}b and \ref{fig_bend_loss}c.
Fig.~\ref{fig_bend_loss}a shows the measured loss of the constant-width TOPIC bends with different $\theta_p$, which agrees well with the theoretical expectations.
When $\theta_p=0^\circ$, the TOPIC bend is a circular bend which has minimal sidewall scattering loss but still shows a high total loss of $0.378\pm0.026$ dB, due to the significant mode mismatch between the straight and circular waveguides.
As $\theta_p$ increases from $0^\circ$ to $22.5^\circ$, the total loss decreases to $0.204\pm0.021$ dB, owing to the smooth transition between the straight and circular modes.
In the $\theta_p$ range of $22.5^\circ$ to $45^\circ$, the loss does not vary much which corresponds to the optimal balance between the sidewall scattering and mode transition losses.
If $\theta_p$ is further increased towards $90^\circ$, the loss increases, because the mode transition loss becomes negligible but the sidewall scattering loss increases due to the longer waveguide length and larger maximum curvature.
Interestingly, the TOPIC bend with $\theta_p=52.5^\circ$ is geometrically similar to an Euler bend, as shown in Fig.~\ref{fig_bend_loss}b.
These two bends have not only similar total lengths, but also similar average curvatures along the waveguides (Fig.~\ref{fig_bend_loss}d).
Moreover, both of them have continuous curvatures along the waveguides, which result in lower losses compared to the circular bend (Fig.~\ref{fig_bend_loss}e).
The main difference between these two bends is that only the TOPIC bend has a continuous curvature derivative, as shown in Fig.~\ref{fig_bend_loss}g, which results in a much lower loss ($0.212\pm0.026$ dB) than the Euler bend ($0.293\pm0.030$ dB).
These results verified that continuous curvature and continuous curvature derivative are essential for low-loss waveguide design as indicated by the theoretical derivation of loss optimization.
Moreover, varying-width TOPIC bends are also fabricated with comparison to the traditional counterpart that formed by circular outer boundary and oval inner boundary~\cite{AdiabaticThermoOptic_OL2013a} as shown in Fig.~\ref{fig_bend_loss}c.
The varying-width TOPIC bend can further reduce the loss dramatically to $0.017\pm0.005$ dB while the traditional counterpart has the lowest loss of $0.242\pm0.006$ dB with $W_{max}=1.4\ \mathrm{\mu m}$ (Fig.~\ref{fig_bend_loss}f).

\begin{figure}[!b]
    \centerline{\includegraphics{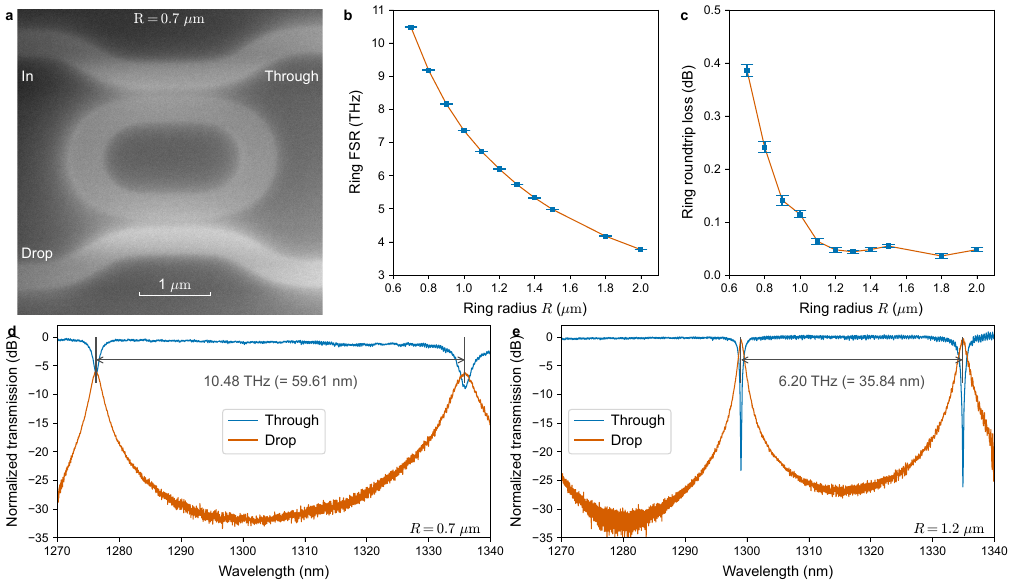}}
    \caption{\textbf{The measured FSR and roundtrip loss of add-drop rings constructed with TOPIC bends.} 
    \textbf{a,} The SEM image of the add-drop ring with radius $R=0.7\ \mathrm{\mu m}$.
    \textbf{b,c,} The measured FSR (\textbf{b}) and roundtrip loss (\textbf{c}) of the rings with different radii.
    \textbf{d,e,} The measured transmission spectra of the rings with radius $R=0.7\ \mathrm{\mu m}$ (\textbf{d}) and $R=1.2\ \mathrm{\mu m}$ (\textbf{e}).
    The nominal waveguide width is 380 nm except the varying-width regions, and the coupling gap is 100 nm.
    All rings take $\theta_{p,i}=43.2^{\circ}$ and $\theta_{p,o}=62.35^{\circ}$ for the TOPIC inner and outer boundaries respectively.
    The ring roundtrip losses are extracted from the measured add-drop spectra at the resonance peak closest to 1300 nm wavelength by the model in Ref.~\cite{LinearRegressionBased_OL2016a}.
    The plotted FSR and roundtrip loss are the mean and standard deviation of the measurements over all dies in a 300 mm wafer.
    The fabrication is done using imec's most advanced iSiPP300 platform allowing high waveguide quality and access to feature dimensions well below 100 nm thanks to the 193 nm immersion lithography.}
    \label{fig_ring_fsr}
\end{figure}

\section{Add-drop ring resonators constructed with TOPIC bends}

\begin{figure}[!b]
    \centering
    \includegraphics{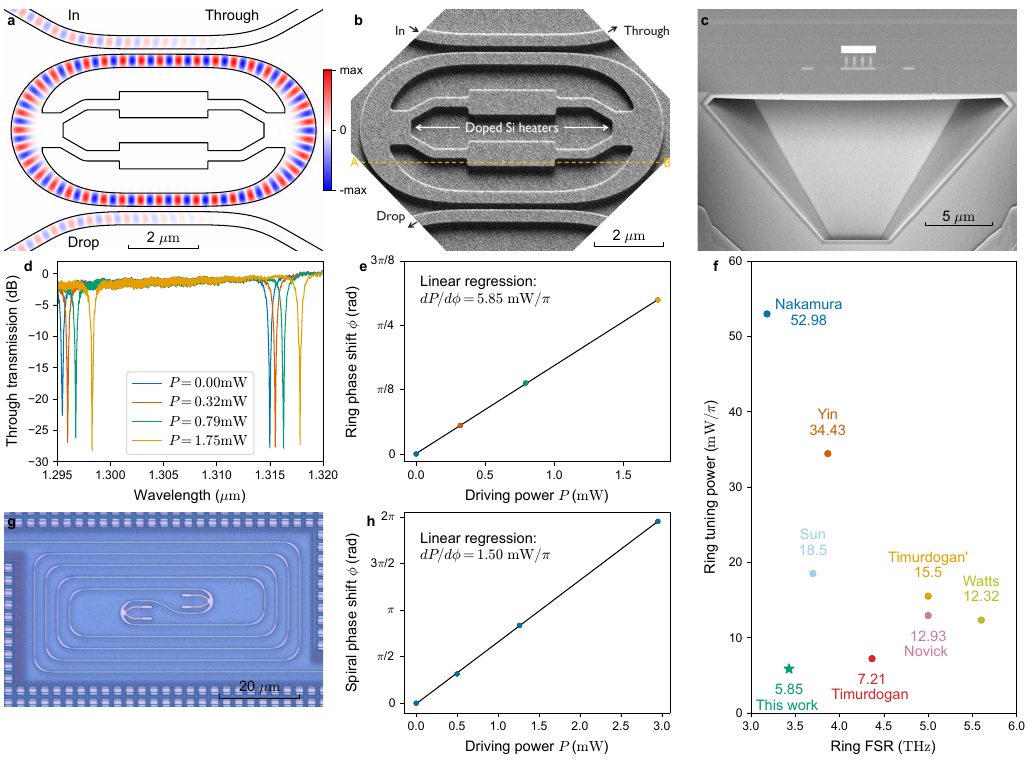}
    \caption{\textbf{The doped silicon heaters embedded in TOPIC bends.}
    \textbf{a,} The optical field evolution ($H$ along thickness direction) at the ring resonance.
    \textbf{b,} The top-view SEM image of the ring before cladding oxide deposition.
    \textbf{c,} The cross-section, corresponding to A-B line in top-view, SEM image of the ring with full material stacks.
    \textbf{d,e,} The measured through spectra under different driving powers (\textbf{d}), and the extracted thermal tuning power (\textbf{e}) of the ring resonator with doped silicon heaters.
    \textbf{f,} The ring tuning power comparison with literature: Nakamura~\cite{HighSpeedChip_AN2022a}, Yin~\cite{SiliconBasedChip_P2023a}, Sun~\cite{IntegratedMicroringTuning_2013a}, Timurdogan'~\cite{AdiabaticResonantMicroring_2013a}, Novick~\cite{LowLosswideFsr_2023a}, Watts~\cite{AdiabaticResonantMicrorings_2COLAEAQEALSCC2V12009a}, and Timurdogan~\cite{LShapedResonant_2013a}.
    \textbf{g,h,} The optical microscope image (\textbf{g}), and the thermal tuning power (\textbf{h}) of the spiral phase shifter with doped silicon heaters.
    The nominal SOI thickness is 220 nm, the waveguide width is 380 nm except the varying-width regions, the coupling gap is 150 nm, and the ring radius $R= 2\ \mathrm{\mu m}$, TOPIC inner boundary transition angle $\theta_{p,i}=43.2^{\circ}$, and TOPIC outer boundary transition angle $\theta_{p,o}=62.35^{\circ}$ in all the rings with doped silicon heaters.
    }
    \label{fig_heater_tuning}
\end{figure}

\begin{figure}[!b]
    \centering
    \includegraphics{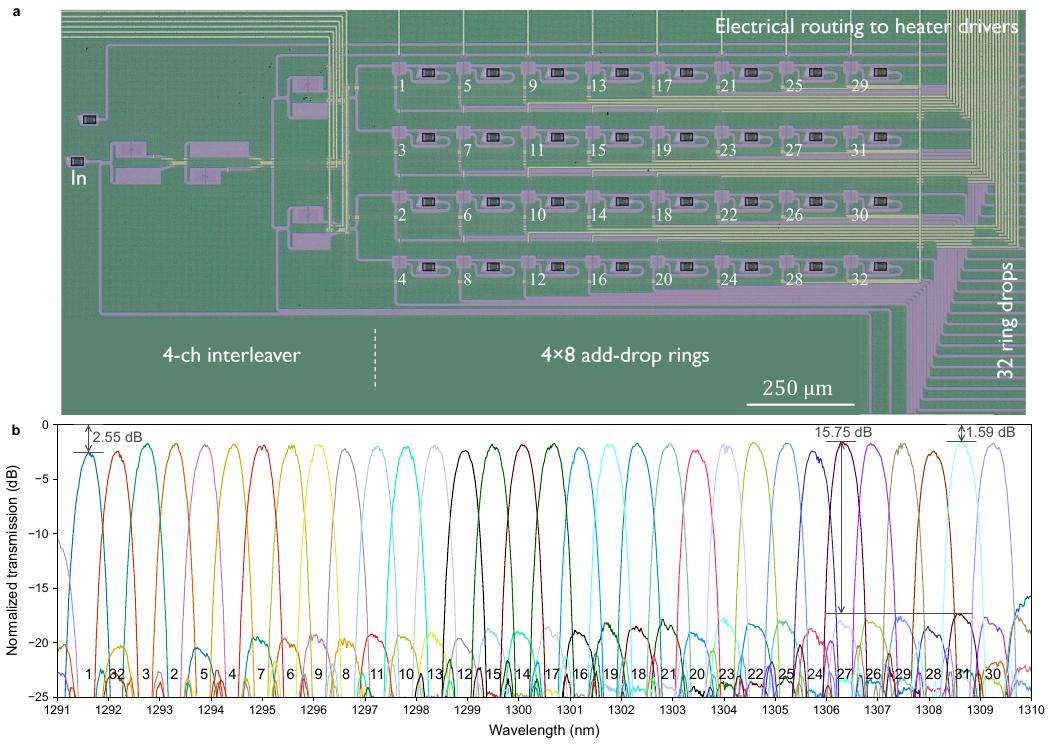}
    \caption{\textbf{The measured spectra of the WDM 32$\mathrm{\times}$100 GHz filter based on TOPIC bends.}
    \textbf{a,b,} The optical microscope image (\textbf{a}) and the measured transmission spectra (\textbf{b}) of the proposed silicon ring-based WDM 32$\mathrm{\times}$100 GHz filter.}
    \label{fig_wdm32x100_spectra}
\end{figure}

A ring resonator can be constructed by two $180^\circ$ TOPIC bends as depicted in Fig.~\ref{fig_ring_fsr}a.
Silicon add-drop rings based on varying-width TOPIC bends with radius $0.7\le R \le 2\ \mathrm{\mu m}$ are fabricated and characterized.
Exploiting the smooth transitions in the TOPIC bend design, the ring with radius of $0.7\ \mathrm{\mu m}$ still shows the typical add-drop ring spectra with only one guided mode excited as shown in Fig.~\ref{fig_ring_fsr}d.
This ring design has achieved an ultra wide FSR of $10.48\pm 0.02$ THz.
However, the ring roundtrip loss is significant due to the substantial radiation loss caused by bending at this ultra-small radius.
With the radius increasing from $0.7\ \mathrm{\mu m}$, the roundtrip loss decreases dramatically (Fig.~\ref{fig_ring_fsr}c), as radiation loss decreases exponentially~\cite{AnalysisCurvedOptical_IJOQE1975a}.
When $R\geq 1.2\ \mathrm{\mu m}$, the TOPIC rings have very low roundtrip loss ($\sim 0.05$ dB) while keeping a wide FSR (Fig.~\ref{fig_ring_fsr}b).
As plotted in Fig.~\ref{fig_ring_fsr}e, the TOPIC ring with $1.2\ \mathrm{\mu m}$ radius shows high-performance add-drop spectra with $6.20 \pm 0.01$ THz FSR, $167\pm 4$ GHz FWHM, $0.42 \pm 0.13$ dB dropping insertion loss, and $28.1 \pm 2.6$ dB dropping extinction ratio reliably in a 300 mm wafer-scale characterization.
Moreover, better performance can be expected when the TOPIC bends are fine optimized.
All rings demonstrated in this paper take $\theta_{p,i}=43.2^{\circ}$ and $\theta_{p,o}=62.35^{\circ}$ for the TOPIC inner and outer boundaries respectively.
These default angle values are set based on initial simulation data without taking the sidewall roughness into account.
However, $\theta_{p,i}$ and $\theta_{p,o}$ should be adjusted to balance the sidewall scattering loss and the mode transition loss in an optimal TOPIC design.
We have implemented this optimization for the radius $2\ \mathrm{\mu m}$, and the measurements show that the ring roundtrip loss is further reduced to $0.032 \pm 0.006$ dB with $\theta_{p,i}=42.5^{\circ}$ and $\theta_{p,o}=65^{\circ}$ while it was $0.047 \pm 0.009$ dB with the default angles.

As indicated by Fig.~\ref{fig_ring_fsr}b, the FSR of the TOPIC rings with radius $1.2\sim2\ \mathrm{\mu m}$ are $\geq 3.2$ THz that is required for WDM 32$\mathrm{\times}$100 GHz.
To implement the design in Fig.~\ref{fig_scheme_3d}a for WDM, the 2 $\mathrm{\mu m}$ radius is chosen since it also provides sufficient space for electrical connections inside the ring.
The 2-$\mathrm{\mu m}$-radius TOPIC ring is used with two 0.81-$\mathrm{\mu m}$-length straight waveguides inserted between the TOPIC bends to achieve $\sim0.09$ ring-bus coupling ratio as required by the 32-channels WDM (Fig.~\ref{fig_scheme_3d}b), while keeping an FSR of 3.423 THz (Fig.~\ref{fig_heater_tuning}d).
Moreover, a whispering gallery mode is excited in the varying-width TOPIC bend so that the light is confined into the waveguide by the outer boundary solely as shown in Fig.~\ref{fig_heater_tuning}a.
The inner boundary of the TOPIC bend serves as an ideal space for integrating doped silicon heaters and connections to the heater driver, as visually depicted in Fig.~\ref{fig_scheme_3d}a and Fig.~\ref{fig_heater_tuning}b.
This meticulous arrangement ensures that the doped silicon and metal wires remain non-overlapping with the light path within the ring, thereby preventing any additional ring optical loss.
Meanwhile, the heating process occurs within the silicon waveguide, enhancing the thermo-optic tuning efficiency owing to the much higher thermal conductivity of silicon than silicon oxide (see Note4, Supplementary Information, for the simulated temperature distribution).
Furthermore, we, at imec, have developed sealed under cut (UCUT) process to remove the substrate silicon under the heaters to further enhance the tunning efficiency (Fig.~\ref{fig_heater_tuning}c).
The resulting ring spectra, obtained under varying driving power levels, are illustrated in Fig.~\ref{fig_heater_tuning}d.
By tracking the resonance wavelength drift in these spectra, the corresponding ring phase shifts are extracted, as graphically represented in Fig.~\ref{fig_heater_tuning}e.
This TOPIC ring yields a remarkable tuning power of 5.85 mW/$\mathrm{\pi}$, surpassing existing literature results for rings with FSR$\geq$3.2 THz (Fig.~\ref{fig_heater_tuning}f).
The FSR$\geq$3.2 THz is emphasized in the comparison because this is required for WDM 32$\mathrm{\times}$100 GHz, and the ring form factor is bounded by the FSR.
Notably, when the ultra-compact form factor is not a must, as in the Mach-Zehnder interferometer (MZI), the tuning power can be further reduced.
For instance, integrating the TOPIC bends with doped silicon heaters into a spiral waveguide, as demonstrated in Fig.~\ref{fig_heater_tuning}g, results in a tuning power of 1.50 mW/$\mathrm{\pi}$ (Fig.~\ref{fig_heater_tuning}h).

\section{Silicon ring-based WDM 32$\mathrm{\times}$100 GHz filter}

In ring-based WDM filters, all the rings coupled to the same bus waveguide will drop signals in a sequential way that degrades the insertion loss and channel isolation of the worst channel accumulatively (see Note5, Supplementary Information, for the measured 8$\times$400 GHz and 16$\times$200 GHz WDM filters based on TOPIC rings without tuning).
To achieve a high-performance 32$\mathrm{\times}$100 GHz WDM filter, the 32 rings are designed to couple with 4 parallel bus waveguides, as illustrated in Fig.~\ref{fig_wdm32x100_spectra}a.
Each bus waveguide is coupled with 8 rings with a channel spacing of 400 GHz.
Across the 4 bus waveguides, the working wavelengths are staggered by 100 GHz.
To divide the 32$\mathrm{\times}$100 GHz input signals into the 4 bus waveguides, the spiral phase shifters are employed to construct a 4-channel cascaded MZI interleaver that divides the input into 4 groups of 8$\mathrm{\times}$400 GHz signals.
The first lattice filter in the cascading is constructed with two phase-shifting stages for flat-top transmission spectra~\cite{CascadedMachZehnder_OE2013a}.
The second phase-shifting stage features an arm relative length that is double that of the first stage.
The two stages are interconnected by three directional couplers, with power cross-coupling ratios of 0.5, 0.29, and 0.08, respectively.
At each phase shifting stage, the phase shifter in the shorter arm is implemented with the doped silicon heaters as shown in Fig.~\ref{fig_heater_tuning}g.
Each output of the interleaver is de-multiplexed by 8 add-drop TOPIC rings (Fig.~\ref{fig_heater_tuning}b) with slight different straight lengths to space the working wavelength.
All the 36 doped silicon heaters in the interleaver and the rings are driven in parallel to align the working wavelengths.
In this way, a WDM 32$\mathrm{\times}$100 GHz filter with fully reconfigurable working wavelengths is achieved.
As a proof of concept, the filter is tuned to the wavelength grid centered at 1300 nm with channel spacing of 100 GHz as defined in the Continuous-Wave Wavelength Division Multiplexing Multi-Source Agreement (CW-WDM MSA)~\cite{CwWdmMsa_CMa}.
The measured spectra are plotted in Fig.~\ref{fig_wdm32x100_spectra}b.
This WDM 32$\mathrm{\times}$100 GHz filter shows channel FWHM of 75$\pm$4 GHz, insertion loss of $1.91\pm 0.28$ dB (1.59 $\sim$ 2.55 dB), channel isolation $\geq 15.75$ dB, and accumulative crosstalk $\leq -15.18$ dB.
The filter tuning efficiency is estimated to be 283 GHz/mW per channel based on the individual characterizations of the ring and the spiral phase shifter.
Compared to the results existing in literature (Tab.~\ref{tab_wdm_comparison}), this filter has doubled the WDM channel amount and achieved much lower insertion loss at the same time.

\section{Conclusion}

Mode mismatch loss between straight and circular waveguides is a well-known problem in silicon photonics.
A lot of bend designs have been explored to reduce the loss by making the curvature transit continuously, such as 
Euler bends~\cite{DramaticSizeReduction_OE2013a,LowLossCompact_OE2017a,LowLossLow_OE2018a,AnalysisSiliconNitride_OE2019a,UltrahighQSilicon_PR2020a},
Bezier bends~\cite{CompactSingleMode_IPJ2011a,MicroringResonatorDesign_2015a,NeuralAdjointMethod_OE2021a,OptimalBezierCurve_OL2021a,LowLossModified_OE2022a},
bends with third order polynomial curvatures~\cite{DesignLowLoss_OAQE2008a,GeneralDesignAlgorithm_OE2012a},
and bends with sinusoidal curvatures~\cite{IdealBendContour_IPTL2007a,UniversalDesignWaveguide_JOLT2019a,GeneralSBend_JOPCS2019a,LowLossWaveguide_JOLT2020a}.
However, these curves are introduced intuitively without a rigorous analysis about the optimal condition for low-loss waveguide design.
In this paper, we have derived the optimal condition with a general optimization methodology, and found that both continuous curvature and continuous curvature derivative are essential to minimize the loss in a waveguide.
Guided by the theoretical derivations, a TOPIC bend is proposed to replace a traditional circular bend with an arbitrary angle.
The TOPIC bend has a unique feature of continuous curvature derivative everywhere compared to all existing low-loss bend designs.
This unique feature allows TOPIC bend to surpass the famous Euler bend for low-loss waveguide designs.
As experimentally demonstrated in this paper, the Euler bend can reduce the loss from $0.378\pm0.026$ dB in a circular bend to $0.293\pm0.030$ dB, while the TOPIC bend achieves $0.212\pm0.026$ dB when it is configured to be geometrically similar to the Euler bend.
Moreover, the TOPIC bend can be easily configured to have varying widths that further reduces the loss to $0.017\pm0.005$ dB, which is 14 times lower than the traditional varying-width bend ($0.242\pm0.006$ dB).
It is worth mentioning that the traditional varying-width bend is implemented with circular outer boundary and oval inner boundary so that it is applicable to $180^\circ$ bends only.
On the contrary, the TOPIC bends can be used with an arbitrary angle, allowing for the features of ultra-compact radius and heater integration.
The TOPIC bend is expected to be widely used to replace all circular bends in all applications of integrated photonics, reducing greatly system size and power consumption.
As a proof of concept, we have achieved ultra-low loss in a bent directional coupler based on the TOPIC bend~\cite{LowLossSilicon_JOLT2024a}.

When the TOPIC bends are deployed into ring resonators, the silicon ring with  a radius as small as $1.2 \ \mathrm{\mu m}$ still exhibits high-quality add drop spectrum thanks to the use of the varying-width TOPIC bends which reduced the ring roundtrip loss to a mere $\sim 0.05$ dB. 
The TOPIC ring with $R=0.7\ \mathrm{\mu m}$ is observed to resonate without the excitation of any high-order mode, which is the smallest radius ever  reported to our best knowledge.
Moreover, the varying-width TOPIC bends can be configured to work with a  WGM as demonstrated with the 2-$\mathrm{\mu m}$-radius TOPIC rings.
The WGM is exploited to integrate a heater into the waveguide bend to improve the thermo-optic tuning efficiency without introducing any extra loss.
The tuning power of 5.85 $\mathrm{mW/\pi}$ is achieved in this paper which makes a new record for silicon rings with FSR$\geq$3.2 THz.
The tuning power is further reduced to 1.50 $\mathrm{mW/\pi}$ by a spiral waveguide design for the applications that do not require an ultra-compact form factor.
With the TOPIC rings and spiral phase shifters, a WDM 32$\times$100 GHz filter is demonstrated with an insertion loss of $1.91\pm 0.28$ dB and a tuning efficiency of 283 GHz/mW per channel.
This filter has doubled the WDM channel amount and achieved much lower insertion loss at the same time, compared to the existing silicon ring-based WDM filters in literature.
These demonstrations prove that the proposed TOPIC bend can unleash the full potential of silicon ring resonators with a large ring FSR, low ring roundtrip loss, scalable ring-bus coupling, and energy-efficient resonance tuning.

\begin{table}[!t]
    \centering
    \caption{\textbf{Performance comparison of silicon ring based WDM filters.}}
    \begin{tabular}{cccccc}
    \hline
        Reference & Channel amount & Channel spacing & Channel FWHM & Insertion loss & Channel isolation \\
        \ & \ & (GHz) & (GHz) & (dB) & (dB) \\ \hline
        \cite{SiliconBasedChip_P2023a} & 16 & 200 & - & 3.9 $\sim$ 5.6 &$\geq 17.1$ \\ 
        \cite{SiMicroRing_OE2011a} & 16 & 100 & 70 & $\sim$5 & $\sim20$ \\ 
        \cite{Novel16Channel_2022a} & 16 & 80 & 32 & - & - \\  
        This work & 32 & 100 & 75$\pm$4 & 1.59 $\sim$ 2.55 & $\geq 15.75$ \\\hline
    \end{tabular}
    \label{tab_wdm_comparison}
\end{table}

\medskip
\textbf{Methods} \par
\textbf{Simulation of the guided modes in circular bends} The eigenmodes in circular bends are simulated by COMSOL mode analysis with the Maxwell equations expressed in a cylindrical coordinate system.

\textbf{UCUT fabrication process} The UCUT process, that is the process to remove the silicon substrate under the SOI, is available in both 200 mm and 300 mm of imec silicon photonics platforms.
The UCUT in this paper is fabricated with the 300 mm platform.
After the cladding oxide deposition on SOI waveguides, a few deep trenches are processed through all oxide layers, cladding oxide and the buried oxide, to the silicon substrate.
These trenches are used as chimneys for UCUT process.
The UCUT cavities are pre-shaped by isotropic dry etching, followed by Tetramethylammonium hydroxide wet etching to form the shape in Fig.~\ref{fig_heater_tuning}c with silicon [111] facets.
Finally the UCUT chimneys are sealed by more deposition of the cladding oxide.

\textbf{Aligning the working wavelength of the WDM filter with CW-WDM MSA grid}
There are 36 output ports in the 32$\times$100 GHz WDM filter (Fig.~\ref{fig_wdm32x100_spectra}a), including the dropping of the 32 rings and the through transmission of the 4 bus waveguides.
These output ports are used for heater tuning with laser inputted from the left In port.
All the 36 heaters are driven in parallel by a programable multiple-channel current source.

Step-1: to align the first lattice filter in the MZI interleaver.
The top 18 outputs are summed up as the monitoring spectrum.
The driving currents of the two heaters in the first lattice filter are tunned till the monitoring spectrum is aligned to CW-WDM MSA grid with maximized spectrum extinction ratio.
The currents are kept at the optimal values for the following steps.

Step-2: to align the top part of the second interleaver cascading.
The top 9 outputs are summed up as the monitoring spectrum.
The driving currents of the corresponding heater is tunned till the monitoring spectrum shows minimum insertion loss.

Step-3: to align the bottom part of the second interleaver cascading.
The 19th$\sim$27th outputs are summed up as the monitoring spectrum.
The driving currents of the corresponding heater is tunned till the monitoring spectrum shows minimum insertion loss.

Step-4: to align the 32 rings one by one.
All the heaters in the MZ interleaver are driven with the optimal currents found in Step 1 to 3.
Then, the rings are aligned one by one, starting from the left.
The driving current of the ring is tuned till the ring drop spectrum appear at the desired wavelength grid with minimum insertion loss.

\textbf{Tuning efficiency estimation of the WDM 32$\times$100 GHz filter} There are 4 spiral phase shifters in the interleaver with a tuning power of 1.50 $\mathrm{mW/\pi}$.
With maximally $2\pi$ of phase shift, the working wavelength of an MZI can be tuned to any specific value.
Therefore, the power consumption of each interleaver phase shifter is estimated to be 3 $\mathrm{mW}$.
The tuning of the interleaver makes all WDM channels shift at the same time such that $1/32$ of the interleaver power consumption is counted for each channel.
Each WDM channel has one ring with a tuning power of 5.85 $\mathrm{mW/\pi}$ and an FSR of $\sim$3.423 THz.
The tuning efficiency of the filter is estimated as $1/(2\times5.85/3423+4\times3/3200/32) =283\ \mathrm{GHz/mW}$ per channel.

\medskip
\textbf{Supporting Information} \par 
Supporting Information is available from the Wiley Online Library or from the author.

\medskip
\textbf{Acknowledgements} \par 
This work was supported in part by imec's industry-affiliation R\&D program on Optical I/O, Research Foundation - Flanders (FWO) under Grant 12ZR720N and PIX4Life (Grant 688519).

\medskip
\textbf{Author contributions} \par 
Q.D. invented the TOPIC bend and designed the devices.
Q.D. wrote the manuscript with reviews from all co-authors.
Q.D., A.HE., H.K., R.M. and J.DC. characterized the devices with Si waveguides.
Q.D. and A.E. finished the simulations.
G.L., C.M. and J.RV. and M.EF. were responsible for the device fabrication and SEM analysis.
P.N. characterized the bends with SiN waveguide.
M.K., Y.T. and K.C. contributed to the doped silicon heater design.
N.S. was responsible for the mask tape-out.
P.VD. managed the research on SiN platform.
M.C., D.V., P.DH., P.V. and P.A. managed the workforce for this project.
J.VC, F.F. and Y.B. managed this project under the research portfolio of imec's industry-affiliation R\&D program on Optical I/O.

\medskip
\textbf{Competing interests} \par 
All authors declare no conflict of interest.
The host institution imec has submitted patent applications (US20240201440A1, EP4386456A1, CN118210104A) for the TOPIC bend.

\medskip


\bibliographystyle{MSP}
\bibliography{wgm_ring_wdm32_ref}

\end{document}


\title[]{Low-Loss and Low-Power Silicon Ring Based WDM 32$\times$100 GHz Filter Enabled by a Novel Bend Design (Supplementary Information)}

\author*[1]{\fnm{Qingzhong} \sur{Deng}}\email{qingzhong.deng@imec.be}
\author[1]{\fnm{Ahmed} \sur{H. El-Saeed}}
\author[1]{\fnm{Alaa} \sur{Elshazly}}
\author[1]{\fnm{Guy} \sur{Lepage}}
\author[1]{\fnm{Chiara} \sur{Marchese}}
\author[1]{\fnm{Pieter} \sur{Neutens}}
\author[1]{\fnm{Hakim} \sur{Kobbi}}
\author[1]{\fnm{Rafal} \sur{Magdziak}}
\author[1]{\fnm{Jeroen} \sur{De Coster}}
\author[1]{\fnm{Javad Rahimi} \sur{Vaskasi}}
\author[1]{\fnm{Minkyu} \sur{Kim}}
\author[1]{\fnm{Yeyu} \sur{Tong}}
\author[1]{\fnm{Neha} \sur{Singh}}
\author[1]{\fnm{Marko} \sur{Ersek Filipcic}}
\author[1]{\fnm{Pol} \sur{Van Dorpe}}
\author[1]{\fnm{Kristof} \sur{Croes}}
\author[1]{\fnm{Maumita} \sur{Chakrabarti}}
\author[1]{\fnm{Dimitrios} \sur{Velenis}}
\author[1]{\fnm{Peter} \sur{De Heyn}}
\author[1]{\fnm{Peter} \sur{Verheyen}}
\author[1]{\fnm{Philippe} \sur{Absil}}
\author[1]{\fnm{Filippo} \sur{Ferraro}}
\author[1]{\fnm{Yoojin} \sur{Ban}}
\author[1]{\fnm{Joris} \sur{Van Campenhout}}

\affil*[1]{\orgname{imec}, \orgaddress{\street{Kapeldreef 75}, \city{Leuven}, \postcode{3001}, \country{Belgium}}}

\maketitle

\section{Theoretical modelling of add-drop ring resonators}

The transmission spectra of a symmetrically coupled add-drop ring resonator has been well established in Ref.~\cite{UniversalRelationsCoupling_EL2000a} as:

\begin{equation}
    \begin{dcases}
    \displaystyle
    &T=\frac{r^2+r^2a^2-2r^2a\cos \theta}{1+r^4a^2-2r^2a\cos \theta}\\
    &D=\frac{a\kappa^4}{1+r^4a^2-2r^2a\cos \theta}\\
    \end{dcases}.
    \label{eq_ring_spectra}
\end{equation}
where $T$ and $D$ are the normalized transmission at the Through and Drop port respectively.
The ring-bus coupling at the input and drop region are identical with $r$ and $\kappa$ as the self- and cross-coupling coefficients respectively.
$\kappa^2$ is the ring-bus coupling ratio of the light power, where the coupling loss is ignored so that $r^2+\kappa^2=1$.
$a$ represents the ring roundtrip amplitude attenuation, which is directly
related to the roundtrip loss of the optical power, $Loss = -10\log_{10}(a^2)$.
The roundtrip phase delay of the ring $\theta=\int_{0}^{L} \frac{2\pi }{\lambda}n_{\mathrm{eff}}ds$ with $n_{\mathrm{eff}}$ as the effective refractive index and $L$ as the geometrical length of the ring.
The maximum dropping power
\begin{equation}
    D_{max} = \frac{a\kappa^4}{(1-r^2a)^2}
\end{equation}
which is used to calculate the dropping insertion loss $IL_{drop}=-10\log_{10}D_{max}$.

According to the the model in Ref.~\cite{LinearRegressionBased_OL2016a}, the ring free spectral range (FSR), in light frequency unit, can be expressed as

\begin{equation}
    \mathrm{FSR}=\frac{c}{OL_g}
\end{equation}
where $c$ is the light speed in vacuum and $OL_g=\int_{0}^{L} (n_{\mathrm{eff}}-\lambda\partial n_{\mathrm{eff}}/\partial \lambda)ds$ is the group length of the ring.
The full width at half maximum (FWHM) of the dropping spectra, in light frequency unit, can be expressed
\begin{equation}
    \mathrm{FWHM}=\frac{2\mathrm{FSR}}{\pi}\arcsin\frac{1-r^2a}{2r\sqrt{a}}
\end{equation}
which is used to calculate the maximum WDM channel amount $Ch_{max}=\mathrm{FSR}/\mathrm{FWHM}$.

\section{Mode mismatch loss calculation in circular bends}
The fundamental TE eigenmode in both the straight and the circular bend are numerically generated by mode analysis in COMSOL.
The difference is that the Maxwell equations are expressed in a cylindrical coordinate system for circular waveguide while a Cartesian coordinate system for straight waveguide.
All the four transverse optical field components are extracted for mode overlapping calculation.
Assuming $+z$ is the light propagation direction, the transverse fields are $\mathbf{E_t}=(E_x,E_y)$ and $\mathbf{H_t}=(H_x,H_y)$.
The mode overlapping ratio of two modes is calculated as
\begin{equation}
    r = \dfrac{|\iint{\mathrm{Re}(\mathbf{E_{t,1}}\times\mathbf{H_{t,2}^*})\mathrm{d}x\mathrm{d}y}|^2}{\iint{\mathrm{Re}(\mathbf{E_{t,1}}\times\mathbf{H_{t,1}^*})\mathrm{d}x\mathrm{d}y} \cdot \iint{\mathrm{Re}(\mathbf{E_{t,2}}\times\mathbf{H_{t,2}^*})\mathrm{d}x\mathrm{d}y}}
\end{equation}
where the subscripts 1, 2 denote the two modes under comparison.
The mode transition loss in Fig. 2c is calculated as $-10\log_{10}(r)$ between the fundamental TE modes in a circular and a straight waveguide.

\section{Derivation of the expressions for TOPIC bends}

The optical loss in a constant-width waveguide can be expressed as
\begin{equation}
    Loss=\int_{\kappa_{1}}^{\kappa_{2}}\left(A+B\kappa+C\frac{\mathrm{d}\kappa}{\mathrm{d}s}+D\left(\frac{\mathrm{d}\kappa}{\mathrm{d}s}\right)^2\right)\frac{\mathrm{d}s}{\mathrm{d}\kappa}d\kappa
    \label{eq_loss_func}
\end{equation}
where $s$ is the distance to the waveguide starting point and the $\kappa$ is the curvature defined along the waveguide central baseline.
To design a low-loss waveguide means, mathematically, to find an expression of $\kappa$ with respect to $s$ that minimizes $Loss$.
This optimization problem can be solved by Euler-Lagrange equation if we define
\begin{equation}
    \mathcal{L}(\kappa,s,s')=As'+ B\kappa s' + C + \frac{D}{s'}
    \label{eq_euler_lagrange_target}
\end{equation}
where $\kappa$ is taken as the variable, $s$ is a function of $\kappa$, and $s'\stackrel{\text{def}}{=}\mathrm{d}s/\mathrm{d}\kappa$.
The optimal solution by Euler-Lagrange equation is expressed as 
\begin{equation}
    \frac{\partial\mathcal{L}}{\partial s} - \frac{\mathrm{d}}{\mathrm{d}\kappa} \frac{\partial\mathcal{L}}{\partial s'}=0
    \label{eq_euler_lagrange_solution_general}.
\end{equation}
Taking Eq.~\ref{eq_euler_lagrange_target} into Eq.~\ref{eq_euler_lagrange_solution_general}, this optimal solution of Eq.~\ref{eq_loss_func} is expressed as 
\begin{equation}
    -\dfrac{\dfrac{\mathrm{d}}{\mathrm{d}\kappa}\dfrac{\mathrm{d}s}{\mathrm{d}\kappa}}{(\dfrac{\mathrm{d}s}{\mathrm{d}\kappa})^3}=\dfrac{B}{2D}= \mathrm{Constant}
    \label{eq_euler_lagrange_solution}.
\end{equation}
To make this solution expressed with $s$ as the variable, the following swapping relationship is noted.
\begin{equation}
    \kappa''\stackrel{\text{def}}{=}\dfrac{\mathrm{d}}{\mathrm{d}s}\dfrac{\mathrm{d}\kappa}{\mathrm{d}s}=\dfrac{\mathrm{d}}{\mathrm{d}s} \dfrac{1}{\mathrm{d}s/\mathrm{d}\kappa} =-\dfrac{\dfrac{\mathrm{d}}{\mathrm{d}\kappa}\dfrac{\mathrm{d}s}{\mathrm{d}\kappa}}{(\dfrac{\mathrm{d}s}{\mathrm{d}\kappa})^3}
    \label{eq_kappa_s_swap}
\end{equation}
With this equation, the optimal solution expressed in Eq.~\ref{eq_euler_lagrange_solution} can be simplified as
\begin{equation} 
    \kappa'' = \mathrm{Constant}
    \label{eq_optimal_solution}.
\end{equation}
This solution indicates that the optical loss is minimized when the second-order curvature derivative ($\kappa''\stackrel{\text{def}}{=}\mathrm{d}^{2}\kappa/\mathrm{d}s^{2}$) has no variation along the waveguide.
The physical meanings of this optimal solution are that the curvature $\kappa$ and the curvature derivative ($\kappa'\stackrel{\text{def}}{=}\mathrm{d}\kappa/\mathrm{d}s$) should be continuous along the waveguide, and the curvature derivative should vary linearly with respect to the waveguide length.
Eq.~\ref{eq_optimal_solution} is equivalent to the general formula of $\kappa$ as 
\begin{equation}
    \kappa=a+bs+cs^2.
    \label{eq_poly2_general_form}
\end{equation}
To derive a low-loss transition between two different waveguides from this equation, the curvature and curvature derivative of the two waveguides, annotated as  ($\kappa_1$, $\kappa'_1$) and ($\kappa_2$, $\kappa'_2$), should be used as the boundary conditions to determine the values of parameter $a$, $b$, and $c$.
Supposing the total transition length is $L$, the boundary conditions can be written as 
\begin{equation}
    \begin{dcases}
        &\kappa|_{s=0}=\kappa_1\\
        &\kappa'|_{s=0}=\kappa'_1\\
        &\kappa|_{s=L}=\kappa_2\\
        &\kappa'|_{s=L}=\kappa'_2\\
    \end{dcases}.
    \label{eq_poly2_boundary}
\end{equation}
With Eq.~\ref{eq_poly2_general_form} and Eq.~\ref{eq_poly2_boundary}, the transition is determined as
\begin{equation}
    \kappa=\kappa_1 + \kappa'_1 s + \dfrac{(\kappa'_2+\kappa'_1)(\kappa'_2-\kappa'_1)}{4(\kappa_2-\kappa_1)}s^2,\ \mathrm{where}\ 0\leq s \leq \dfrac{2(\kappa_2-\kappa_1)}{\kappa'_2+\kappa'_1}
    \label{eq_poly2_solution}
\end{equation}
which is meaningful only when
\begin{equation} 
    (\kappa_2-\kappa_1)(\kappa'_2+\kappa'_1)\neq 0
    \label{eq_validation_condition}.
\end{equation}
However, the transition between a straight and a circular waveguide does not meet the condition given by Eq.~\ref{eq_validation_condition}.
Thus, we relaxed the optimal solution from ``no variation in $\kappa''$'' to ``no abrupt variation in $\kappa''$''.
In particular, a linear variation of $\kappa''$ with respect to $s$ is used as the relaxed optimal solution that results in the general formula of $\kappa$ as 
\begin{equation}
    \kappa=a+bs+cs^2+ds^3.
    \label{eq_poly3_general_form}
\end{equation}
Supposing the total transition length is $L$, the boundary conditions to transit from straight to circular can be written as 
\begin{equation}
    \begin{dcases}
        &\kappa|_{s=0}=0\\
        &\kappa'|_{s=0}=0\\
        &\kappa|_{s=L}=1/R_c\\
        &\kappa'|_{s=L}=0\\
    \end{dcases}.
    \label{eq_poly3_boundary}
\end{equation}
where $R_c$ is the radius of the circular bend. 
The curvature can be expressed as
\begin{equation}
    \kappa=\dfrac{3L s^2-2s^3}{R_{c}L^{3}}
    \label{eq_poly3_solution_no_angle}.
\end{equation}
According to Eq.~\ref{eq_poly3_solution_no_angle}, a straight-circular transition can be generated when the total transition length $L$ is specified.
We choose to use the total transition angle $\theta_p$ to control the length for practical convenience, which mathematically is
\begin{equation}
    \int_{0}^{L}\kappa \mathrm{d}s=\theta_p.
    \label{eq_poly3_boundary_angle}
\end{equation}
The straight-circular transition is expressed as
\begin{equation}
    \kappa=\dfrac{3R_{c}\theta_{p} s^2-s^3}{4R_{c}^{4}\theta_{p}^{3}},\ \mathrm{where}\ 0\leq s \leq 2R_{c}\theta_{p}.
    \label{eq_top_curvature}
\end{equation}

\section{Simulations of TOPIC bends with a doped silicon heater}
\begin{figure}[!b]%
    \centering
    \includegraphics{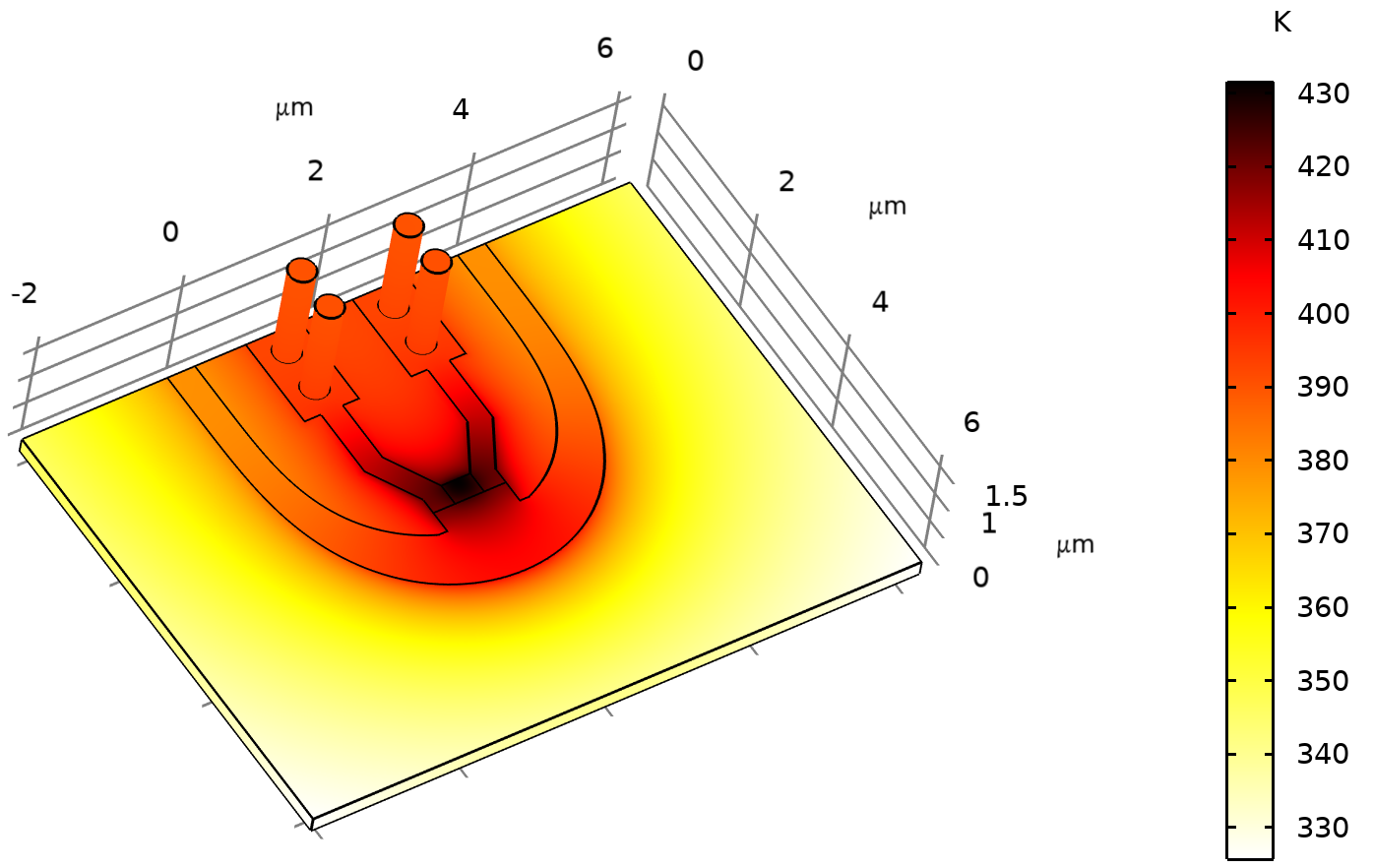}
    \caption{\textbf{The simulated temperature distribution in the TOPIC bend.}
    The materials are the same as described in Fig.1a, where intrinsic Silicon is used as the optical waveguide, N-doped Silicon is used as the heater embedded in the inner boundary of the waveguide, N$^+$-doped Silicon is used to connect the heaters with the Tungsten vias for electrical driving.
    The driving voltage is 2 V.
    The SOI thickness is 210 nm, the bend radius $R= 2\ \mathrm{\mu m}$, the TOPIC inner and outer boundary transition angle are $\theta_{p,i}=43.2^{\circ}$ and $\theta_{p,o}=62.35^{\circ}$ respectively.
    The structure is surrounded by silicon oxide.
    A rectangular air cavity is placed under the 2 $\mathrm{\mu m}$ bottom oxide to mimic the UCUT.
    All the top oxide, bottom oxide and air cavity are included in the simulation, but hidden in this figure to enable the temperature display at the waveguide surface.
    }
    \label{sfig_temperature_distribution}
\end{figure}

The sheet resistances of N- and N$^+$- doped Silicon materials are set to 1335 and 52.4 $\Omega/sq$ respectively based on the measurements with the applied doping processes.
The driving voltage is set to 2 V, resulting in 0.4 mA of current passing through the heater.
As indicated by the temperature distribution in Fig.~\ref{sfig_temperature_distribution}, the heating is mainly concentrated in the silicon region because silicon material has a thermal conductivity of 130 $W/(m\cdot K)$, that is much larger than 1.38 $W/(m\cdot K)$ in the surrounding silicon oxide.

\section{TOPIC ring based WDM-8$\times$400 GHz}
\begin{figure}[!ht]%
    \centering
    \includegraphics{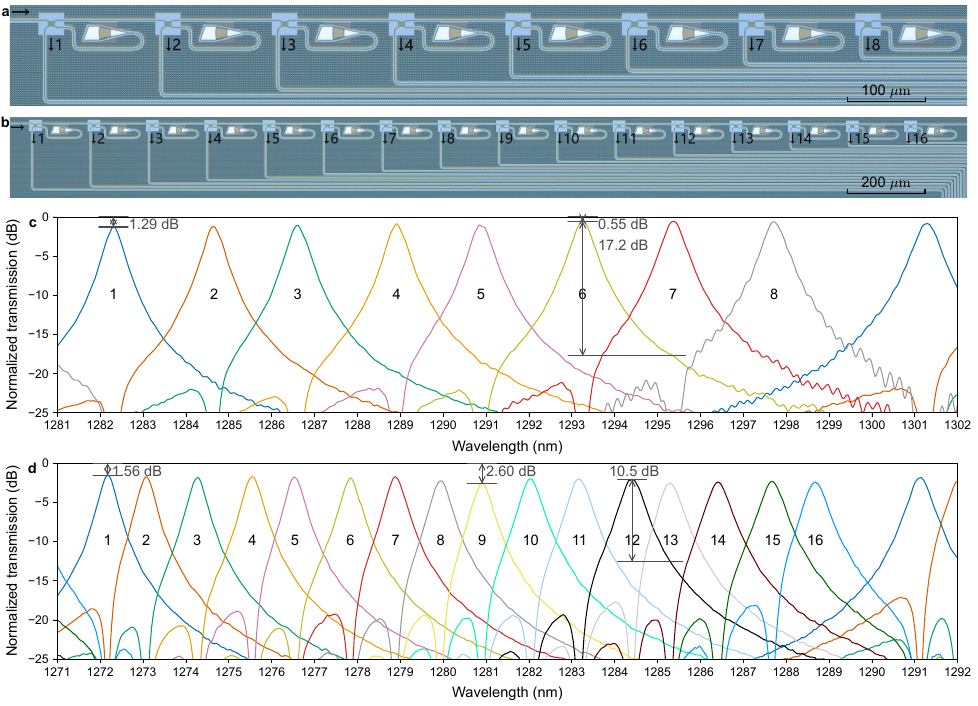}
    \caption{\textbf{The TOPIC ring based WDM filters without tuning.}
    The microscope images (\textbf{a, b}) and the corresponding measured transmission spectra (\textbf{c, d}) of the TOPIC ring based 8$\times$400 GHz (\textbf{a, c}) and 16$\times$200 GHz (\textbf{b, d}) WDM filters.
    The nominal SOI thickness is 220 nm, the waveguide width is 380 nm except the varying-width regions, the coupling gap is 150 nm, and the ring radius $R= 2\ \mathrm{\mu m}$, TOPIC inner boundary transition angle $\theta_{p,i}=43.2^{\circ}$, and TOPIC outer boundary transition angle $\theta_{p,o}=62.35^{\circ}$ in all the rings.
    The nominal straight segment length in the coupling regions are varied in the range of $0.898\sim 1.128$ and $0.754\sim 0.996$ $\mathrm{\mu m}$ to achieve the desired channel spacing, while the nominal ring-bus coupling gap is 158 nm and 150 nm to ahcieve $\sim 0.09$ power cross-coupling in the 8$\times$400 GHz and 16$\times$200 GHz WDM filters respectively.
    }
    \label{sfig_passive_wdm}
\end{figure}

The TOPIC rings are configured as a 8$\times$400 GHz (Fig.~\ref{sfig_passive_wdm}a) and a 16$\times$200 GHz (Fig.~\ref{sfig_passive_wdm}b) WDM filter without tuning.
As indicated by the measured spectra in Fig.~\ref{sfig_passive_wdm}c, the WDM 8$\times$400 GHz filter has very good performance with insertion loss (IL) of $0.91\pm 0.29$ (0.55$\sim$1.29) dB and channel isolation (CI) of $\geq 17.2$ dB.
The performance is degraded to $\mathrm{IL} = 2.06\pm 0.32$ (1.56$\sim$2.60) dB and $\mathrm{CI} \geq 10.5$ dB when it is configured as a WDM 16$\times$200 GHz filter as shown in Fig.~\ref{sfig_passive_wdm}d.
The reasons is that a larger ring count requires a smaller channel spacing under a certain FSR which decreases the CI because the ring transmission spectra are Lorentzian-shaped.
Moreover, all the rings drop signals from the bus waveguide sequentially so that the IL is accumulatively degraded at the same time.
\bibliography{../main_latex/wgm_ring_wdm32_ref}